\documentclass[10pt,conference]{IEEEtran}

\usepackage{amsmath}
\usepackage{amssymb}
\usepackage{graphicx}
\usepackage[table]{xcolor}
\usepackage[vlined]{algorithm2e}
\usepackage{mathpartir}
\usepackage{stmaryrd}
\usepackage[font=footnotesize]{caption} \DeclareCaptionType{copyrightbox}

\usepackage{verbatimbox}
\usepackage{subcaption}
\usepackage[T1]{fontenc}

\usepackage[breaklinks,colorlinks=true]{hyperref}
\usepackage{stackengine}

\newenvironment{proof}{\begin{IEEEproof}}{\end{IEEEproof}\ignorespacesafterend}
\newcommand{\st}{\cdot}

\newtheorem{theorem}{{\bf Theorem}}

\newcommand{\arr}[2]{\texttt{arr}({#1},{#2})}
\renewcommand{\int}{\texttt{int}\xspace}
\newcommand{\bool}{\texttt{bool}\xspace}
\newcommand{\rd}[2]{\mathit{rd}({#1},{#2})}
\renewcommand{\wr}[3]{\mathit{wr}({#1},{#2},{#3})}
\newcommand{\peq}[3]{{#1} =_{#3} {#2}}
\renewcommand{\vec}[1]{\overline{#1}}
\newcommand{\bigand}{\bigwedge}
\newcommand{\bigor}{\bigvee}
\newcommand{\idx}{\vec{i}}
\newcommand{\val}{\vec{v}}
\newcommand{\sig}{\mathcal{S}}
\newcommand{\preds}{\mathcal{P}}

\definecolor{midgrey}{rgb}{0.3,0.3,0.3}
\definecolor{darkred}{rgb}{0.7,0.1,0.1}

\newcommand{\ak}[1]{}
\newcommand{\ag}[1]{}
\newcommand{\nsb}[1]{}

\newcommand{\spacer}{\textsc{Spacer}\xspace}
\newcommand{\seahorn}{\textsc{SeaHorn}\xspace}

\newcommand{\proj}[1]{\mathit{Proj}_{#1}}
\newcommand{\qf}[1]{{#1}_{\mathit{qf}}}

\newcommand{\Inv}{\mathit{Inv}}
\newcommand{\Init}{\mathit{init}}
\newcommand{\Tr}{\mathit{tr}}
\newcommand{\Bad}{\mathit{bad}}
\newcommand{\xcall}{\vec{x}^{o}}
\newcommand{\cF}{\mathcal{F}}

\newcommand{\may}{\mathcal{O}}
\newcommand{\must}{\mathcal{U}}

\newcommand{\Queue}{\mathit{Q}}

\newcommand{\nat}{\mathbb{N}}

\newcommand{\MBP}{\textsc{Mbp}\xspace}
\newcommand{\Itp}{\textsc{Itp}\xspace}

\makeatletter
\newcommand{\removelatexerror}{\let\@latex@error\@gobble}
\makeatother

\IEEEoverridecommandlockouts

\begin{document}
% needed to get \thanks to work
\IEEEoverridecommandlockouts

%\title{Quantifier-free Solutions for Horn Clauses over Integers and Arrays}
\title{Compositional Verification of Procedural Programs using Horn Clauses over Integers and Arrays}

\newcommand{\seimarkings}{This material is based upon work funded and supported by the
Department of Defense under Contract No. FA8721-05-C-0003 with
Carnegie Mellon University for the operation of the Software
Engineering Institute, a federally funded research and development
center. This material has been approved for public release and
unlimited distribution. DM-0002442.}

\author{
    \IEEEauthorblockN{Anvesh Komuravelli}
    \IEEEauthorblockA{Computer Science Department \\ Carnegie Mellon University \\
        Pittsburgh, PA, USA}
\and
    \IEEEauthorblockN{Nikolaj Bj{\o}rner}
    \IEEEauthorblockA{Microsoft Research \\ Redmond, WA, USA}
\and
    \IEEEauthorblockN{Arie Gurfinkel\thanks{\tiny\seimarkings}}
    \IEEEauthorblockA{Software Engineering Institute \\ Carnegie Mellon University \\
        Pittsburgh, PA, USA}
\and
    \IEEEauthorblockN{Kenneth L. McMillan}
    \IEEEauthorblockA{Microsoft Research \\ Redmond, WA, USA}
}

\maketitle

\begin{abstract}
We present a compositional SMT-based algorithm for safety of procedural C
programs that takes the heap into consideration as well. Existing SMT-based
approaches are either largely restricted to handling linear arithmetic
operations and properties, or are non-compositional. We use Constrained Horn
Clauses (CHCs) to represent the verification conditions where the memory
operations are modeled using the extensional theory of arrays (ARR). First, we
describe an exponential time quantifier elimination (QE) algorithm for ARR which
can introduce new quantifiers of the index and value sorts. Second, we adapt the
QE algorithm to efficiently obtain under-approximations using models, resulting
in a polynomial time Model Based Projection (MBP) algorithm. Third, we integrate
the MBP algorithm into the framework of compositional reasoning of procedural
programs using may and must summaries recently proposed by us. Our solutions to
the CHCs are currently restricted to quantifier-free formulas. Finally,
we describe our practical experience over SV-COMP'15 benchmarks using an
implementation in the tool \spacer.
\end{abstract}

%\begin{IEEEkeywords}
%model checking, smt, quantifier elimination, model, compositional reasoning,
%memory, arrays, horn clauses, summary
%\end{IEEEkeywords}

\section{Introduction}
\label{sec:intro}

Under-approximating a projection (i.e., existential quantification), for example
in computing an image, is a key aspect of many techniques of symbolic model
checking. A typical (though not ubiquitous) approach to this is what we call
\emph{Model-based Projection} (MBP)~\cite{spacer_cav14}: we generalize a particular
point in the space of the image (obtained using a model) to a subset of the
image that contains it.  In some cases, the purpose is to compute the exact
image by a series of under-approximations~\cite{gupta}. In other cases, such as
IC3~\cite{ic3}, the purpose of MBP is to produce a relevant proof sub-goal.  When
the number of possible generalizations is finite, we say that we have a \emph{finite
MBP} which allows us to compute the exact image by iterative sampling, or to
guarantee that the branching in our proof search is finite.

The feasibility of a finite MBP depends on the underlying logical theory.
Finite MBPs exist for propositional logic~\cite{gupta,gpdr} and Linear
Integer Arithmetic (LIA) with a divisibility predicate~\cite{spacer_cav14}, and have been applied in both
hardware and software model checking. LIA is often adequate
for software verification, provided that heap
and array accesses can be eliminated. This can be done by abstraction,
or by inlining all procedures and performing compiler optimizations to
lower memory into registers (e.g.,~\cite{ufo,seahorn_svcomp15}).  However, the
inlining approach has many drawbacks. It can expand the
program size exponentially, it cannot handle recursion, and it is not always
feasible to eliminate heap and array accesses.
%cannot handle unbounded arrays or heap allocation.

We address this issue here by considering the problem of MBP for the
extensional theory of arrays (ARR). We find that a finite MBP exists that can be
computed in polynomial time when only array-valued variables are
projected. Projecting variables of index and value sorts is not
always possible, since the quantifier-free fragments of the theory
combinations are not guaranteed to be closed under projection. We therefore take a pragmatic approach to
MBP that may not always converge to the exact projection. This allows us to
handle, for example, the combination of ARR and LIA.

We test the effectiveness of this approach using the model checking framework of
\spacer~\cite{spacer_cav14}. This SMT-based framework makes use of MBP
to produce proof sub-goals for Hoare-style procedure-modular proofs
of recursive programs.  The ability to reason with ARR makes
it possible to handle heap-allocating programs without inlining
procedures, as the heap can be faithfully modeled using ARR~\cite{seahorn}. This leads to significant
improvements in scalability, when compared to the use of LIA
alone with inlining, as measured using benchmark programs from the 2015 Software
Verification Competition (SVCOMP 2015)~\cite{svcomp15}. Not
inlining the programs also has the advantage that we generate
procedure-modular proofs (containing procedure summaries) that might
be re-usable in various ways (e.g.,~\cite{evolcheck}).

In summary, we (a)~describe an exponential rewriting procedure for projecting
array variables (Sec.~\ref{sec:qe}), (b)~adapt this procedure to obtain a
polynomial-time (per model) finite MBP for projecting array variables
(Sec.~\ref{sec:mbp}), (c)~integrate this with existing MBP procedures for Linear
Arithmetic (Sec.~\ref{sec:arr_lia}) in the
\spacer framework obtaining a new compositional proof search algorithm
(Sec.~\ref{sec:framework}), and (d)~evaluate the algorithm experimentally using
SVCOMP benchmarks (Sec.~\ref{sec:results}).

%This paper is organized as follows. In Sec.~\ref{sec:qe} we define the
%notion of model-based projection. We describe an exponential rewriting
%procedure for projecting array variables, then show how this can be
%adapted to a polynomial procedure (per model) for MBP. This in turn
%yields a model generalization procedure for arbitrary projections. In
%Sec.~\ref{sec:qfree} we describe the integration of this approach with
%existing MBP procedures for Linear Arithmetic in the \spacer framework
%to obtain a new compositional proof search algorithm
%(Sec.~\ref{sec:qfree}). In Sec.~\ref{sec:results} we the approach experimentally.

%%% Local Variables:
%%% mode: latex
%%% TeX-master: "main"
%%% End:

\section{Preliminaries}

We consider a first-order language with equality whose signature $\sig$ contains basic sorts (e.g.,
\bool of Booleans, \int of integers, etc.) and array sorts. An
array sort $\arr{I}{V}$ is parameterized by a sort of indices $I$ and a
sort of values $V$. We assume that $I$ is always a basic sort. For every array
sort $\arr{I}{V}$, the language has the usual function symbols
$\mathit{rd} : \arr{I}{V} \times I \to V$ and $\mathit{wr} :
\arr{I}{V} \times I \times V \to \arr{I}{V}$ for reading from
and writing to the array. Intuitively, $\rd{a}{i}$ denotes the value stored in
the array $a$ at the index $i$ and $\wr{a}{i}{v}$ denotes the array obtained
from $a$ by replacing the value at the index $i$ by $v$. We use the following
axioms for the extensional theory of arrays (ARR):

\textbf{Read-after-write}\\
\indent $\forall a : \arr{I}{V} ~\forall i,j : I ~\forall v : V$
\begin{align*}
& \left( i=j \implies \rd{\wr{a}{i}{v}}{j} = v \right) \land \\
& \left( i \neq j \implies \rd{\wr{a}{i}{v}}{j} = \rd{a}{j} \right)
\end{align*}
\indent\textbf{Extensionality}\\
\indent $\forall a,b : \arr{I}{V} \st \left( \forall i : I \st \rd{a}{i} = \rd{b}{i} \right) \implies a=b$

Intuitively, the first schema says that after modifying an array $a$ at index
$i$, a read results in the new value at index $i$ and $\rd{a}{j}$ at every
other index $j$. The second schema says that if two arrays agree on the values
at every index location, the arrays are equal. We use an over-bar to denote a vector.
We write $\vec{x}: S$ to denote that every term in vector $\vec{x}$ has sort
$S$, $\vec{x}(k)$ to denote the $k$th component of $\vec{x}$, and $y \in
\vec{x}$ to denote that $y$ is equal to some component of $\vec{x}$, i.e.,
$\bigor_{k=1}^{|\vec{x}|} y = \vec{x}(k)$.
%The notation $\val : V$ indicates that every term in vector $\val$ has sort $V$, while
%$w \in \val$ indicates that $w$ is equal to some component of $\val$.
Let $\idx : I$ and
$\val : V$ be vectors of index and value terms of the same length
$m$. We write $\wr{a}{\idx}{\val}$ to denote
$\wr{\mathit{wr}(\dots\wr{a}{\idx(0)}{\val(0)}\dots)}{\idx(m)}{\val(m)}$.
%where $\vec{i}(k)$ denotes the $k$th element of $\vec{i}$.  For an index term $j : I$,
%we write $j \in \idx$ to denote $\bigor_{k=1}^m (j = \idx(k))$.
%
Unless specified otherwise, $\sig$ contains no other symbols.

%\emph{Partial Equalities}.
For arrays $a$ and $b$ of sort $\arr{I}{V}$, and a (possibly empty) vector of
index terms $\idx$, we write $\peq{a}{b}{\idx}$ to denote $\forall j : I \st
\left( j \not\in \idx \implies \rd{a}{j} = \rd{b}{j} \right)$ and call such formulas
\emph{partial equalities}~\cite{stump}. Using extensionality, one can easily
show the following
\begin{align}
  \peq{a}{b}{\emptyset} &\equiv  a=b                    \label{eq:peq_intro} \\
  \peq{\wr{a}{j}{v}}{b}{\idx} &\equiv
        \begin{aligned}
            & \left( j \in \idx \land \peq{a}{b}{\idx} \right) \lor \\
            & \left( j \not\in \idx \land \peq{a}{b}{\idx,j} \land \rd{b}{j} = v \right)
        \end{aligned}                                 \label{eq:peq} \\
  \peq{a}{b}{\idx} &\equiv
        \exists \val : V \st a = \wr{b}{\idx}{\val}  \label{eq:peq2}
\end{align}

We write $\varphi(\vec{x})$ for a formula $\varphi$ with free
variables $\vec{x}$, and we treat $\phi$ as a predicate over
$\vec{x}$.  We also write $\varphi[t]$ to to indicate that a term or
formula $t$ occurs in $\varphi$ at some syntactic position.

Given formulas $\varphi_A(\vec{x}, \vec{z})$ and $\varphi_B(\vec{y},\vec{z})$
with $\vec{x} \cap \vec{y} = \emptyset$ and $\varphi_A
\implies \varphi_B$, a Craig Interpolant~\cite{craig}, denoted $\Itp(\varphi_A,
\varphi_B)$, is a formula
$\varphi_I(\vec{z})$ such that $\varphi_A \implies \varphi_I$ and $\varphi_I
\implies \varphi_B$.

\section{QE and MBP for the theory ARR}

\begin{figure*}[t]
\begin{mathpar}
    \inferrule*[left=ElimWrRd,rightskip=-1cm]
    {\varphi[\rd{\wr{t}{i}{v}}{j}]}
    {(i=j \land \varphi[v]) \lor
     (i \neq j \land \varphi[\rd{t}{j}])}
    \inferrule*[left=ElimWrEq]
    {\varphi[\peq{\wr{t_1}{j}{v}}{t_2}{\idx}]}
    {
      \left( j \in \idx \land \varphi[\peq{t_1}{t_2}{\idx}] \right) ~\lor\\\\
     \left( j \not\in \idx \land \varphi[\peq{t_1}{t_2}{\idx,j} \land v = \rd{t_2}{j}] \right)
    }
\end{mathpar}

\begin{mathpar}
    \inferrule*[left=PartialEq,right=\normalfont\textrm{$t_i$'s have array sort}]
    {\varphi[t_1=t_2]}
    {\varphi[\peq{t_1}{t_2}{\emptyset}]}

    \inferrule*[left=TrivEq]
    {\varphi[\peq{t}{t}{\idx}]}
    {\varphi[\top]}

    \inferrule*[left=Symm,right=\normalfont\textrm{\def\stackalignment{l}\stackanchor{$t_2$
    is a write term}{but $t_1$ is not}}]
    {\varphi[\peq{t_1}{t_2}{\idx}]}
    {\varphi[\peq{t_2}{t_1}{\idx}]}
\end{mathpar}

\vspace{.5cm}
\begin{center}
\small
\textsc{ElimWr = (ElimWrRd $\mid$ ElimWrEq $\mid$ PartialEq $\mid$ TrivEq $\mid$ Symm)}
\end{center}
\caption{Rewriting rules to eliminate write terms. \textsc{ElimWr} denotes
one of the rules chosen non-deterministically.}
\label{fig:str_rules}
\end{figure*}

\begin{figure*}[t]
\begin{mathpar}
    \inferrule*[left=CaseSplitEq,rightskip=-.5cm]
    {\exists a \st \varphi[\peq{a}{t}{\idx}]}
    {\exists a \st \left( \left( \peq{a}{t}{\idx} \land \varphi[\top] \right)
     \lor
     \left( \neg (\peq{a}{t}{\idx}) \land \varphi[\bot] \right) \right)}
    \inferrule*[left=FactorRd,right=\normalfont\textrm{\def\stackalignment{l}\stackanchor{$s$
is fresh, $t$ does not}{contain array terms}}]
    {\exists a \st \varphi[\rd{a}{t}]}
    {\exists a, s \st (\varphi[s] \land s=\rd{a}{t})}
\end{mathpar}
\caption{Rewriting rules to factor out equalities and read terms on the quantified
array variable.}
\label{fig:eq_sel_rules}
\end{figure*}

By projection of a variable we mean elimination of an existential
quantifier.  Consider a formula $\varphi$ of the form $\exists \vec{x}
\st \qf{\varphi}(\vec{x}, \vec{y})$ where $\qf{\varphi}$ is
quantifier-free. The problem of \emph{quantifier elimination} (QE) in
$\varphi$ is to find a logically equivalent quantifier-free formula
$\psi(\vec{y})$. In this case, we say that $\psi$ is the result of
projecting $\vec{x}$ in $\qf{\varphi}$.

A \emph{model-based projection} (MBP) for $\varphi$ is an operator
$\proj{}$ that takes a model $M$ of $\qf{\varphi}$ and returns a
quantifier-free formula $\psi_M(\vec{y})$ such that $M \models \psi_M$
and $\psi_M$ entails $\varphi$. The operator $\proj{}$ is a
\emph{finite MBP} if its image is finite up to logical
equivalence (that is, over all models we obtain only finitely many
semantically distinct formulas).\footnote{
MBP as defined in~\cite{spacer_cav14} corresponds to \emph{finite MBP} here.
} In this case, we obtain the exact
projection as the disjunction of the image of $\proj{}$. We will
refer to $\proj{}(M)$ as a \emph{generalization} of $M$.

In some cases, there is a trivial approach to MBP that we
will call the \emph{substitution approach}. We simply substitute for each
variable $x$ in $\varphi$ a constant that is equal to $x$ in the given model
$M$ (for example, a numeric literal). This approach was taken for
propositional logic by Ganai et al.~\cite{gupta}. For theories that admit models of
unbounded size (e.g., LIA), however, this does not
yield a finite MBP, as the number of distinct generalizations we obtain can be
infinite.
%\footnote{In the case of real linear arithmetic, though, a
  %finite number of ``virtual'' substitutions suffice, if we allow
  %infinitessimal values and
  %$\pm\infty$~\cite{LoosWeisspfenning}.}

Instead, we can take the approach used for Linear Real Arithmetic and LIA in our
earlier work~\cite{spacer_cav14}. Suppose that for the given theory we have a
QE procedure that produces a formula with an
exponential (or higher) number of disjunctions. We can adapt this
procedure to an MBP by always choosing just one disjunct that is true
in the given model $M$. The result may be a procedure that is polynomial for any
given model, though the number of distinct generalizations is
exponential. We will show how to apply this idea for the projection of
array-valued variables in the theory of arrays ARR. When combining
this theory with LIA, we will find that some variables of index and value sorts
must be eliminated by the substitution method, which gives us a useful
MBP but not necessarily a \emph{finite} MBP.

\subsection{Quantifier elimination for ARR}
\label{sec:qe}

Consider an existentially quantified formula $\exists a : \arr{I}{V} \st \varphi$
where $\varphi$ is quantifier-free. While we cannot always obtain an equivalent
quantifier-free formula, our objective here is to obtain an equivalent
existentially quantified formula where every quantifier (if any) is of the sort $I$ or $V$.
As a simplification, we restrict
the interpretations of $I$, the index sort, to infinite domains. Handling finite
index domains requires a slight adaptation of the algorithms as described in
Appendix~\ref{app:finite_dom}.

\begin{figure}[t]
\centering

\begin{mathpar}
    \inferrule*[left=ElimEq]
    {\exists a \st (\peq{a}{t}{\idx} \land \varphi)}
    {\exists \val \st \varphi[\wr{t}{\idx}{\val}/a]}
\end{mathpar}
\flushright{where $a$ does not appear in $t$ and $\val$ denotes fresh variables}

\begin{mathpar}
    \inferrule*[left=ElimDiseq]
    {\exists a \st \left( \varphi \land \bigand_{k=1}^{m} \neg (\peq{a}{t_k}{\idx_k}) \right)}
    {\exists a \st \varphi}
\end{mathpar}
\flushright{where $m \in \nat$, $a$ does not appear in any $t_k$, and\\
$a$ appears in $\varphi$ only in read terms over $a$}

\begin{mathpar}
    \inferrule*[left=Ackermann]
    {\exists a \st \left( \varphi \land \bigand_{k=1}^{m} s_k = \rd{a}{t_k} \right)}
    {\varphi \land \bigand_{1 \le k < \ell \le m} \left( t_k=t_\ell \implies s_k=s_\ell \right) }
\end{mathpar}
\flushright{where $m \in \nat$ and $a$ does not appear in $\varphi$, $s_k$'s, or $t_k$'s}

\caption{Rewriting rules for QE of arrays.}
\label{fig:core_qe_rules}
\end{figure}

\begin{algorithm}
\small
\DontPrintSemicolon
\SetKwBlock{Begin}{\textmd{\textsc{ArrayQE}}$(\exists a \st \varphi)$}{}
\Begin{
    \nl $\varphi_1 \gets (\textsc{ElimWr}^*)(\exists a \st \varphi)$
    \nl $\varphi_2 \gets (\textsc{CaseSplitEq}^* ; \textsc{FactorRd}^*)(\varphi_1)$\;
    \nl $\left( \bigor_{k=1}^n \delta_k \right) \gets \textsc{LiftEqDiseqRd}(\varphi_2)$\;
    \nl \For{$k \in [1,n]$} {
    \nl     $\psi_k \gets (\textsc{ElimEq} ; \textsc{ElimDiseq} ; \textsc{Ackermann})(\delta_k)$
        }
    \nl \Return $\bigor_{k=1}^n \psi_k$
}
\caption{QE for $\exists a \st \varphi$, where $a$ is an array variable.}
\label{alg:qe}
\end{algorithm}

Our algorithm is inspired by the decision procedure for the quantifier-free fragment
of ARR by Stump \emph{et al}.~\cite{stump}. At a high level, the QE algorithm proceeds
in 3 steps: (i) eliminate write terms using the read-after-write axiom schema
and partial equalities over arrays, (ii) eliminate (partial) equalities and
disequalities over arrays, and (iii) eliminate read terms over arrays.
Alg.~\ref{alg:qe} shows the pseudo-code for our QE algorithm \textsc{ArrayQE}
using the rewrite rules in Fig.~\ref{fig:str_rules}, \ref{fig:eq_sel_rules},
and~\ref{fig:core_qe_rules}. Each rule rewrites the formula above the line to the
logically equivalent formula below the line. We use regular expression notation to
express sequences of rewrites. In particular, Kleene star applied to a rule denotes
the rule's application to a fixed point.

Line~1 of \textsc{ArrayQE} eliminates write terms using the rewrite
rules in Fig.~\ref{fig:str_rules}. Here \textsc{ElimWr} denotes a rule
in Fig.~\ref{fig:str_rules} chosen non-deterministically.
\textsc{ElimWrRd} rewrites terms using the read-after-write axiom and
\textsc{ElimWrEq} rewrites partial equalities using
Eq.~(\ref{eq:peq}). \textsc{PartialEq} converts equalities into
partial equalities using Eq.~(\ref{eq:peq_intro}). \textsc{TrivEq}
eliminates trivial partial equalities with identical arguments and
\textsc{Symm} ensures that write terms on the r.h.s. of equalities are
also eliminated.

Line~2 of \textsc{ArrayQE} rewrites the formula by case-splitting on partial
equalities on the array quantifier $a$ (via \textsc{CaseSplitEq}) followed by
factoring out read terms over $a$ by introducing new quantifiers of sort $V$
(via \textsc{FactorRd}). Note that, as presented, these two rules are not terminating as the partial equalities and read terms are preserved in
the conclusion of the rules. However, one can easily ensure that a given partial
equality or read term is considered exactly once by first computing the set
of all partial equalities and read terms in the formula and processing them in
a sequential order. The details are straightforward and are left to the reader.

\textsc{LiftEqDiseqRd} on line~3 of \textsc{ArrayQE} performs Boolean rewriting
and returns an equivalent disjunction such that in every disjunct, the partial
equalities, array disequalities, and equalities over read terms appear at the
end as conjuncts, in that order.
%In practice, \textsc{CaseSplitEq} can be
%implemented efficiently using an $(N+1)$-way case analysis for a total number of $N$
%partial equalities in the formula and this Boolean rewriting can be
%avoided.
For each disjunct, line~5 applies the rules in Fig.~\ref{fig:core_qe_rules} to
eliminate the array quantifier $a$.  \textsc{ElimEq} obtains a substitution term
for $a$ using the equivalence in Eq.~(\ref{eq:peq2}). \textsc{ElimDiseq} is
applicable when the disjunct contains no partial equalities and given that the
domain of interpretation of $I$ is infinite, one can always satisfy the
disequalities and hence, they can simply be dropped. \textsc{Ackermann} performs
the Ackermann reduction~\cite{ackermann} to eliminate the read terms.

Note that while the rewrite rules are applicable to all array terms and
equalities in the original formula, in practice, we only need to apply them to
eliminate the relevant terms containing the array quantifier $a$.
See Fig.~\ref{fig:qe_eg} for an illustration of \textsc{ArrayQE} on an example.

\begin{figure*}[t]
\small
$\exists a \st \left( b = \wr{a}{i_1}{v_1} \lor (\rd{\wr{a}{i_2}{v_2}}{i_3} > 5 \land \rd{a}{i_4} > 0) \right)$
\begin{align*}
\equiv~ & \exists a \st
        \begin{aligned}
        & \left( \boldsymbol{i_2 = i_3} \land (b = \wr{a}{i_1}{v_1} \lor (\boldsymbol{v_2 > 5} \land \rd{a}{i_4} > 0)) \right) \lor \\
        & \left( \boldsymbol{i_2 \neq i_3} \land (b = \wr{a}{i_1}{v_1} \lor (\boldsymbol{\rd{a}{i_3} > 5} \land \rd{a}{i_4} > 0)) \right) \\
        \end{aligned} & \{\textsc{ElimWrRd}\} \\
\equiv~ & \exists a \st
        \begin{aligned}
        & \left( i_2 = i_3 \land (\boldsymbol{(\peq{a}{b}{i_1} \land \rd{b}{i_1} = v_1)} \lor (v_2 > 5 \land \rd{a}{i_4} > 0)) \right) \lor \\
        & \left( i_2 \neq i_3 \land (\boldsymbol{(\peq{a}{b}{i_1} \land \rd{b}{i_1} = v_1)} \lor (\rd{a}{i_3} > 5 \land \rd{a}{i_4} > 0)) \right) \\
        \end{aligned} & \{\textsc{PartialEq};\textsc{ElimWrEq}\} \\
\equiv~ & \exists a \st
        \begin{aligned}
        & \left( \boldsymbol{\peq{a}{b}{i_1}} \land
            \begin{aligned}
                & \left( i_2 = i_3 \land (\rd{b}{i_1} = v_1 \lor (v_2 > 5 \land \rd{a}{i_4} > 0)) \right) \lor \\
                & \left( i_2 \neq i_3 \land (\rd{b}{i_1} = v_1 \lor (\rd{a}{i_3} > 5 \land \rd{a}{i_4} > 0)) \right)
            \end{aligned} \right) ~\lor \\
        & \left( \boldsymbol{\neg (\peq{a}{b}{i_1})} \land
            \begin{aligned}
                & \left( i_2 = i_3 \land (v_2 > 5 \land \rd{a}{i_4} > 0) \right) \lor \\
                & \left( i_2 \neq i_3 \land (\rd{a}{i_3} > 5 \land \rd{a}{i_4} > 0) \right)
            \end{aligned} \right) \\
        \end{aligned} & \{\textsc{CaseSplitEq}\} \\
\equiv~ & \exists a,s_3,s_4 \st
        \begin{aligned}
        & \Biggl( \left( \peq{a}{b}{i_1} \land
            \underbrace{
            \begin{aligned}
                & \left( i_2 = i_3 \land (\rd{b}{i_1} = v_1 \lor (v_2 > 5 \land \boldsymbol{s_4} > 0)) \right) \lor \\
                & \left( i_2 \neq i_3 \land (\rd{b}{i_1} = v_1 \lor (\boldsymbol{s_3 > 5} \land \boldsymbol{s_4} > 0)) \right)
            \end{aligned}
            }_{\varphi_1}
            \right) ~\lor \\
        & \left( \neg (\peq{a}{b}{i_1}) \land
            \underbrace{
            \begin{aligned}
                & \left( i_2 = i_3 \land (v_2 > 5 \land \boldsymbol{s_4} > 0) \right) \lor \\
                & \left( i_2 \neq i_3 \land (\boldsymbol{s_3 > 5} \land \boldsymbol{s_4} > 0) \right)
            \end{aligned}
            }_{\varphi_2}
            \right) \Biggr) \\
        &  \land \boldsymbol{s_3 = \rd{a}{i_3} \land s_4 = \rd{a}{i_4}} \\
        \end{aligned} & \{\textsc{FactorRd}\} \\
\equiv~ & \exists a,s_3,s_4 \st
        \begin{aligned}
        & \left( \varphi_1 \land \boldsymbol{\peq{a}{b}{i_1} \land s_3 = \rd{a}{i_3} \land s_4 = \rd{a}{i_4}} \right) ~\lor \\
        & \left( \varphi_2 \land \boldsymbol{\neg (\peq{a}{b}{i_1}) \land s_3 = \rd{a}{i_3} \land s_4 = \rd{a}{i_4}} \right) \\
        \end{aligned} & \{\textsc{LiftEqDiseqRd}\} \\
\equiv~ & \exists v,s_3,s_4 \st \left( \varphi_1 \land s_3 = \rd{a}{i_3} \land s_4 = \rd{a}{i_4} \right)\boldsymbol{[\wr{b}{i_1}{v}/a]} ~\lor & \{\textsc{ElimEq}\}\\
        & \exists a,s_3,s_4 \st \left( \varphi_2 \land s_3 = \rd{a}{i_3} \land s_4 = \rd{a}{i_4} \right) & \{\textsc{ElimDiseq}\}\\
\equiv~ & \exists v,s_3,s_4 \st \left( \varphi_1 \land s_3 = \rd{a}{i_3} \land s_4 = \rd{a}{i_4} \right)[\wr{b}{i_1}{v}/a] ~\lor & \\
        & \exists s_3,s_4 \st \left( \varphi_2 \land \boldsymbol{(i_3 = i_4 \implies s_3 = s_4)} \right) & \{\textsc{Ackermann}\}\\
\end{align*}
\caption{Illustrating \textsc{ArrayQE} on an example.}
\label{fig:qe_eg}
\end{figure*}

\textbf{Correctness and Complexity}.
We can show the following properties of \textsc{ArrayQE}
(proof sketches in Appendix~\ref{app:qe_proofs}).

\begin{theorem}
    $\textsc{ArrayQE}(\exists a : \arr{I}{V} \st \varphi)$ returns $\exists \val
    : V \st \rho$, where $\rho$ is quantifier-free and $\exists \val \st \rho
    \equiv \exists a \st \varphi$.
\end{theorem}

\begin{theorem}
    $\textsc{ArrayQE}(\exists a \st \varphi)$ terminates in time exponential in
    the size of $\varphi$.
\end{theorem}

\subsection{Model Based Projection}
\label{sec:mbp}

In this section, we will assume that for a satisfiable formula we can
obtain a finite representation of a model of the formula and that we
can effectively evaluate the truth of any formula in this model. This is
possible for ARR and its combinations with LIA and propositional logic. The
ability to evaluate allows us to strengthen a formula in a way that
preserves a given model. Suppose we have a formula
$\varphi[\psi_1\vee\psi_2]$ with model $M$, where the sub-formula
$\psi_1\vee\psi_2$ occurs positively (under an even number of
negations) in $\varphi$.  If we also have $M \models \psi_1$, then  $M \models
\varphi[\psi_1]$ and clearly, $\varphi[\psi_1]$ entails
$\varphi$. This gives us a way to eliminate a disjunction while
preserving a given model and maintaining an under-approximation.
If neither $\psi_1$ nor $\psi_2$ is true in $M$,
we can similarly replace $\varphi$ with $\varphi[\bot]$. These transformations
are expressed as MBP rewrite rules in Fig.~\ref{fig:mbp_rules}. 

For each QE rule $R$, we can produce a corresponding
under-approximate rule $R_M$ that preserves model $M$. This rule can be
written $R~;~(\textsc{MbpLeft} \mid \textsc{MbpRight} \mid
\textsc{MbpVac})^*$. In practice, we can choose to only apply the MBP rules to 
disjunctions introduced by the QE rules and not to those originally occurring in $\varphi$.
Correspondingly, we can convert our QE algorithm
$\textsc{ArrayQE}$ to $\textsc{ArrayQE}_M$ by replacing each rule
$R$ with $R_M$. We can then obtain an MBP $\textsc{ArrayMBP}(\varphi)(M) =
\textsc{ArrayQE}_M(\varphi)$ and we can show the following:

\begin{theorem}
    For any quantifier-free formula $\varphi$ in ARR, $\textsc{ArrayMBP}(\exists a : \arr{I}{V}.\ \varphi)$ is
    a finite MBP.
\end{theorem}

The fact that it is an MBP can be easily shown by induction on the number of
rewrites applied.  The fact that it is finite derives from the fact
that there are only finitely many ways to resolve the disjunctions in
the QE result.

Moreover, assuming that the evaluation of a formula in a model can be done in
polynomial time, we can evaluate $\textsc{ArrayMBP}(\varphi)(M)$ in time
that is polynomial in the size of $M$ and the size of~$\varphi$. This is because
we can polynomially bound the number of times each rule $R_M$ applies, and
each rule can only expand the formula size by a constant amount. Fig.~\ref{fig:mbp_eg}
shows an example of applying $\textsc{ArrayMBP}$.

\begin{figure}[t]
\centering

\begin{mathpar}
    \inferrule*[left=MbpLeft]
    {\varphi[\psi_1\vee \psi_2] \\ M \models \varphi, \psi_1}
    {\varphi[\psi_1] }

    \inferrule*[left=MbpRight]
    {\varphi[\psi_1\vee \psi_2] \\ M \models \varphi, \psi_2}
    {\varphi[\psi_2] }

    \inferrule*[left=MbpVac]
    {\varphi[\psi_1\vee \psi_2] \\ M \models \varphi \\ M \not\models \psi_1,\psi_2}
    {\varphi[\bot] }
\end{mathpar}

\caption{MBP rules for formulas in negation-normal form.}
\label{fig:mbp_rules}
\end{figure}

\begin{figure*}[t]
\small
$\exists a \st \left( b = \wr{a}{i_1}{v_1} \lor (\rd{\wr{a}{i_2}{v_2}}{i_3} > 5 \land \rd{a}{i_4} > 0) \right)$
\begin{align*}
\Leftarrow~ & \exists a \st
        \left( \boldsymbol{i_2 \neq i_3} \land (b = \wr{a}{i_1}{v_1} \lor (\boldsymbol{\rd{a}{i_3} > 5} \land \rd{a}{i_4} > 0)) \right)
        & \{\textsc{WrRd}_M, M \models i_2 \neq i_3\} \\
\Leftarrow~ & \exists a \st
        \left( i_2 \neq i_3 \land (\boldsymbol{(\peq{a}{b}{i_1} \land \rd{b}{i_1} = v_1)} \lor (\rd{a}{i_3} > 5 \land \rd{a}{i_4} > 0)) \right)
        & \{\textsc{PartialEq};\textsc{WrEq}_M\} \\
\Leftarrow~ & \exists a \st
            \boldsymbol{\neg (\peq{a}{b}{i_1})} \land i_2 \neq i_3 \land (\rd{a}{i_3} > 5 \land \rd{a}{i_4} > 0)
            & \{\textsc{CaseEq}_M, M \not\models \peq{a}{b}{i_1}\} \\
\Leftarrow~ & \exists a,s_3,s_4 \st
        \begin{aligned}
            & \left( \neg (\peq{a}{b}{i_1}) \land
            \underbrace{
                i_2 \neq i_3 \land (\boldsymbol{s_3 > 5} \land \boldsymbol{s_4} > 0)
            }_{\varphi_2}
            \right) \\
            & \land \boldsymbol{s_3 = \rd{a}{i_3} \land s_4 = \rd{a}{i_4}}
        \end{aligned}
        & \{\textsc{FactorRd}\} \\
\Leftarrow~ & \exists a,s_3,s_4 \st
        \left( \varphi_2 \land \boldsymbol{\neg (\peq{a}{b}{i_1}) \land s_3 = \rd{a}{i_3} \land s_4 = \rd{a}{i_4}} \right)
        & \{\textsc{LiftEqDiseqRd}\} \\
\Leftarrow~ & \exists a,s_3,s_4 \st \left( \varphi_2 \land s_3 = \rd{a}{i_3} \land s_4 = \rd{a}{i_4} \right) & \{\textsc{ElimDiseq}\}\\
\Leftarrow~ & \exists s_3,s_4 \st \left( \varphi_2 \land \boldsymbol{(i_3 = i_4 \land s_3 =
s_4)} \right) & \{\textsc{Ack}_M, M \models i_3=i_4\}\\
\end{align*}
\caption{Illustrating \textsc{ArrayMBP} on the example of Fig.~\ref{fig:qe_eg} with a given model $M$.}
\label{fig:mbp_eg}
\end{figure*}

\subsection{MBP for ARR+LIA}
\label{sec:arr_lia}

%\ak{we seem to say that MBP is in our hands, Itp is not. That is, we are asking
%how can spacer terminate for an arbitrary Itp procedure, under what restrictions
%on the CHCs, etc.? Correct?}

We now consider the combination of the ARR and LIA theories.
Assume that the only basic sorts are \bool and \int. Furthermore, we only consider linear
functions over \int along with a divisibility predicate (with constant divisors).
We developed a finite MBP for LIA in a previous work~\cite{spacer_cav14} (call
it \textsc{LiaMBP}). When the index sort $I$ is $\int$, one can obtain a more
efficient MBP with a slight modification of $\textsc{Ackermann}_M$ (for
eliminating array read terms) that utilizes the predicate symbol $<$. Given a
model $M$ of the formula, one can first partition the set of index terms $t_k$'s
according to their interpretations in $M$ and choose a representative for each
equivalence class. Then, the conjunction in the result of the rule is modified
as follows: (a) for every equivalence class, add the equality $t_k = t_\ell$
for every non-representative $t_\ell$, where $t_k$ is the representative, (b) linearly
order the representatives and add the corresponding inequalities. The modified
rule (and hence, the resulting MBP) is linear in time and space.

%When the index sort $I$ is $\int$, one can obtain a modified
%$\textsc{Ackermann}_M$ rule (for eliminating array read terms) by utilizing the predicate
%symbol $<$ and the given model $M$ to linearly order the index terms $t_k$'s.
%This can be achieved by partitioning the set of index terms $t_k$'s according to
%their interpretations in $M$, choosing a representative for each equivalence
%class, ensuring that $t_k$ is always a representative in the equality $t_k =
%t_\ell$ in the rule, and linearly ordering the representatives of the various
%equivalence classes according to $M$. The resulting MBP is linear in time and space and
%is more efficient (at the price of an enlarged image set).

However, the combination of arrays and integers introduces terms over the
combined signature which need to be handled as well. For example, there is no
equivalent quantifier-free formula for $\exists i : \int \st \rd{a}{i} > 0$.
This implies that there does not exist a finite MBP for the combination of LIA and
ARR. In the example, the only way to under-approximate the quantification is to
use the substitution method, replacing $i$ with its interpretation in a model $M \models \rd{a}{i} > 0$
as a numeric literal.

Based on the above observations, we obtain an MBP for ARR+LIA as follows.
First, we apply \textsc{ArrayMBP}, using the modified $\textsc{Ackermann}_M$
above, to eliminate array quantifiers. Then, we use \textsc{LiaMBP} to
eliminate integer quantifiers that do not appear in any array term. Finally, we
use the substitution method to eliminate any remaining integer quantifiers. When
the last step of substitution method is not necessary, the resulting MBP will be finite.

\section{The Compositional Verification Framework}
\label{sec:framework}

MBP plays a crucial role in enabling the search for compositional proofs. In
this section, we will consider the role played by MBP in a model
checking framework called \spacer~\cite{spacer_cav14}.  In this framework, MBP is
used to create succinct localized proof sub-goals that make it
possible to reason about only one procedure at a time. The proof goals
take the form of under-approximate summaries, either of the calling
context of a procedure or of the procedure itself. Without some form
of projection, \spacer would not be compositional, as it would build up
formulas of exponential size, in effect inlining procedures to create
bounded model checking formulas.

\subsection{Modeling programs with CHCs}

\spacer checks safety of procedural programs by reducing the problem to SMT of a special kind of
formulas known as \emph{Constrained Horn Clauses} (CHCs)~\cite{bmr12,spacer_cav14,seahorn}.
%\emph{Constrained Horn Clauses}.
We augment the signature $\sig$ with a set of fresh predicate symbols $\preds$.
A \emph{Constrained Horn Clause} (CHC) is a formula of the form
\[
\forall \vec{x} \st
    \underbrace{\bigand_{k=1}^m P_k(\vec{x}_k) \land \varphi(\vec{x})}_{\mathit{body}}
    \implies
    \mathit{head}
\]
where for each $k$, $P_k$ is a symbol in $\preds$, $\vec{x}_k \subseteq \vec{x}$ and $|\vec{x}_k|$ is equal to the arity of $P_k$. The constraint $\varphi$ is a formula over $\sig$, and $\mathit{head}$ is
either an application of a predicate in $\preds$ or another formula over $\sig$.
We use \emph{body} to refer to the antecedent of the CHC,
as shown above.  A CHC is called a \emph{query} if $\mathit{head}$ is a
formula over $\sig$ and otherwise, it is called a \emph{rule}.
% definition of a fact -- I don't see it being used anywhere.
If $m \le 1$ in the body, the CHC is \emph{linear} and is
\emph{non-linear} otherwise. Following the convention of logic programming
literature, we also write the above CHC as $\mathit{head} \gets P_1(\vec{x}_1),
\dots, P_m(\vec{x}_m), \varphi(\vec{x})$.

Intuitively, each predicate symbol $P_k$ represents an unknown partial
correctness specification of a procedure (that is, an over-approximate
summary). A query defines a property to be proved, while each rule
gives modular verification condition for one procedure. A satisfying
assignment to the symbols $P_k$ is thus a certificate that the program
satisfies its specification and corresponds to the annotations in a
Floyd/Hoare style proof. In this work, we are interested in finding
annotations that can be expressed in the \emph{quantifier-free}
fragment of our first-order language, to avoid the difficulty of
reasoning with quantifiers.

Any given set of CHCs encoding safety of procedural programs can be transformed to an equisatisfiable set of
just three CHCs with a single predicate symbol (encoding the program
location using a variable). These CHCs have the following form:

\begin{equation}
\begin{gathered}
  \Inv(\vec{x}) \gets \Init(\vec{x}) \qquad
  \neg\Bad(\vec{x}) \gets \Inv(\vec{x}) \\
  \Inv(\vec{x}') \gets \Inv(\vec{x}), \Inv(\xcall), \Tr(\vec{x}, \xcall, \vec{x}')
\end{gathered}
\label{eq:safety}
\end{equation}

Intuitively, $\Inv$ is the program invariant, $\vec{x}$ denotes the
pre-state of a program transition, $\vec{x}'$ denotes the post-state,
and $\xcall$ denotes the summary of a procedure call (if one is made).
If there are no procedure calls, $\Tr$ is independent of $\xcall$ and
$\Inv(\xcall)$ can be dropped: in this case $\Inv$ denotes an
inductive invariant of an ordinary transition system. In the sequel,
we restrict to this normal form and consider only quantifier-free
interpretations of the predicate $\Inv$.

It is useful to rewrite the above rules using a function $\cF$ that
substitutes given predicates $\phi_A(\vec{x})$ and $\phi_B(\vec{x})$
for the occurrences of $\Inv$ in the rule bodies. That is, let
\[
\begin{split}
\cF(\varphi_A, \varphi_B) \equiv & \left( \varphi_A(\vec{x}) \land
\varphi_B(\xcall) \land \Tr(\vec{x}, \xcall, \vec{x}') \right) \\
& \lor \Init(\vec{x}') 
\end{split}
\]
The rules are thus equivalent to $\cF(\Inv,\Inv) \Rightarrow \Inv(\vec{x})$.
Abusing notation, we will also write $\cF(\varphi_A)$ for $\cF(\varphi_A,
\varphi_A)$.

\subsection{The \spacer framework}

\spacer\ is a general framework that can be instantiated for a given
logical theory $T$ by supplying three elements: \emph{(a)} a model-generating
SMT solver for $T$, \emph{(b)} an MBP procedure \MBP for $T$ and
\emph{(c)} in interpolation procedure \textsc{Itp} for $T$.  Compared to
other SMT-based algorithms (e.g.,~\cite{whale,hsf,ultimate,duality}),
the key distinguishing feature of \spacer is compositional
reasoning. That is, instead of checking satisfiability of large
formulas generated by program unwinding, \spacer iteratively creates
and checks local reachability queries for individual procedures.  In
this way it is similar to IC3~\cite{ic3,pdr}, a SAT-based algorithm
for safety of finite-state transition systems, and GPDR~\cite{gpdr},
its extension to Linear Real Arithmetic. Like these methods, \spacer
maintains a sequence of over-approximations of procedure behaviors,
called \emph{may summaries}, corresponding to program
unwindings. However, unlike other approaches, \spacer also maintains
under-approximations of procedure behaviors, called \emph{must
  summaries}, to avoid redundant reachability queries. Another
distinguishing feature of \spacer is the use of MBP for efficiently
handling existentially quantified formulas to create a new query or a
must summary.  We note, however, that MBP is a general technique and
can be exploited in IC3/PDR as well.\footnote{Arguably sub-goal
  creation in IC3 is a simple MBP for propositional logic.}

Alg.~\ref{alg:spacer} gives a simplified description of \spacer as a solver for
CHCs in the form of (\ref{eq:safety}) (though \spacer\ handles general CHCs).
It is described using a set of rules that can be applied non-deterministically.
Each rule is presented as a guarded command ``[ \emph{grd} ] \emph{cmd}'', where
\emph{cmd} can be executed only if \emph{grd} holds. 
%We will briefly describe the rules below and then mention
%some implementation aspects.

\begin{algorithm}
\DontPrintSemicolon
%\SetKwBlock{Algo}{\textmd{\textsc{Spacer}}$(\Init(\vec{x}), \Tr(\vec{x},\xcall,\vec{x}'), \Bad(\vec{x}))$}{}
\SetKwFor{Forever}{forever}{do}{endfor}
\SetKwIF{Guard}{Guard1}{Guard2}{[}{]}{[}{[}{endguard}
\SetKwComment{Rule}{}{}
%\Algo{
\KwIn{Formulas $\Init(\vec{x}), \Tr(\vec{x}, \xcall, \vec{x}'), \Bad(\vec{x})$}
  
\KwOut{Inductive invariant (FO interpretation of $\Inv$ satisfying
(\ref{eq:safety})) or \textsc{Unsafe}}
\;
\lIf{$(\Init \land \Bad)$ satisfiable} {\Return \textsc{Unsafe}}
\SetCommentSty{emph}
  \tcp*[h]{initialize data structures}\;
  $\Queue := \emptyset$ \tcp*{set of pairs $\langle \varphi, i \rangle, i \in \nat$}
  $N := 0$ \tcp*{max level, or recursion depth}
  $\may_0 = \Init, \may_i = \top$, $\forall i > 0$ \tcp*{may summary sequence}
  $\must = \Init$ \tcp*{must summary}
\SetCommentSty{textbf}
  \Forever{non-deterministically}{
\AlgoDontDisplayBlockMarkers\SetAlgoNoEnd\SetAlgoNoLine
    \Rule*[h]{(Candidate)}
    \Guard {$(\may_N \land \Bad)$ satisfiable} {
        $\Queue := \Queue \cup \langle \varphi, N \rangle$, for some $\varphi \implies \may_N \land \Bad$
    }
    \Rule*[h]{(DecideMust)}
    \Guard
        {$\langle \varphi, i+1 \rangle \in \Queue$,
            $M \models \cF(\may_i, \must) \land \varphi'$} {
            $\Queue := \Queue \cup
                        \langle \MBP(\exists \xcall, \vec{x}' \st \cF(\may_i, \must) \land \varphi', M), i \rangle$
        }
    \Rule*[h]{(DecideMay)}
    \Guard
        {$(\varphi, i+1) \in \Queue$,
            $M \models \cF(\may_i) \land \varphi'$} {
            $\Queue := \Queue \cup
                        \langle \MBP(\exists \vec{x}, \vec{x}' \st \cF(\may_i) \land \varphi', M)[\vec{x}/\xcall], i \rangle$
        }
    \Rule*[h]{(Leaf)}
    \Guard
        {$(\varphi, i) \in \Queue$, $\cF(\may_{i-1}) \implies \neg \varphi'$, $i < N$} {
            $\Queue := \Queue \cup \langle \varphi, i+1 \rangle$
        }
    \Rule*[h]{(Successor)}
    \Guard
        {$\langle \varphi, i+1 \rangle \in \Queue$,
            $M \models \cF(\must) \land \varphi'$} {
                $\must := \must \lor \MBP(\exists \vec{x}, \xcall \st \cF(\must) \land \varphi', M)[\vec{x}/\vec{x}']$
        }
    \Rule*[h]{(Conflict)}
    \Guard
        {$\langle \varphi, i+1 \rangle \in \Queue$,
            $\cF(\may_i) \implies \neg\varphi'$} {
                $\may_j := \may_j \land \Itp(\cF(\may_i),
            \neg\varphi')[\vec{x}/\vec{x}']$, $\forall j \le i+1$
        }
    \Rule*[h]{(Induction)} 
    \Guard
        {$(\varphi \lor \psi) \in \may_i$, $\cF(\varphi \land \may_i) \implies \varphi'$} {
                $\may_j := \may_j \land \varphi$, $\forall j \le i+1$
        }
    \Rule*[h]{(Unfold)}
    \lGuard {$\may_N \implies \neg\Bad$} {
        $N := N + 1$
    }
    \Rule*[h]{(Safe)}
    \lGuard {$\may_{i+1} \implies \may_i$} {
        \Return invariant $\may_i$
    }
    \Rule*[h]{(Unsafe)}
    \lGuard {$(\must \land \Bad)$ satisfiable} {
        \Return \textsc{Unsafe}
    }
  }  
  %}
  \caption{Rule-based description of \spacer.}
  \label{alg:spacer}
\end{algorithm}

As shown in Alg.~\ref{alg:spacer}, \spacer maintains a set of reachability
queries $\Queue$, a sequence of may summaries $\{\may_i\}_{i \in \nat}$, and a
must summary $\must$. Intuitively, a query $\langle \varphi, i \rangle$
corresponds to checking if $\varphi$ is reachable for recursion depth $i$,
$\may_i$ over-approximates the reachable states for recursion depth $i$, and
$\must$ under-approximates the reachable states. $N$ denotes the current bound
on recursion depth. The sequence of may summaries and $N$ correspond to the
\emph{trace of approximations} and the maximum \emph{level} in IC3/PDR,
respectively. For convenience, let $\may_{-1}$ be $\bot$. $\MBP(\varphi, M)$,
for a formula $\varphi = \exists \vec{v} \st \qf{\varphi}$ and model $M \models
\qf{\varphi}$, denotes the result of some MBP function associated with
$\varphi$ for the model $M$.

Alg.~\ref{alg:spacer} initializes $N$ to 0 and, $\may_0$ and $\must$ to $\Init$.
%Then, it iteratively applies one of the rules in the \textbf{forever} loop, chosen non-deterministically.
\textbf{Candidate} initiates a backward search for a
counterexample beginning with a set of states in $\Bad$. The potential
counterexample is expanded using either \textbf{DecideMust} or
\textbf{DecideMay}. \textbf{DecideMust} \emph{jumps over} the call
$\Inv(\xcall)$, in the last CHC of (\ref{eq:safety}), utilizing the must summary
$\must$. \textbf{DecideMay}, on the other hand, creates a query for the call
using the may summary of its calling context.
%\textbf{Leaf} moves an unreachable query to a higher recursion depth.
\textbf{Successor} updates $\must$ when a query is known to be reachable.
%\textbf{Conflict} updates may summaries when a query is known to be
%unreachable. \textbf{Induction} strengthens may summaries using induction
%relative to $\may_i$. \ak{comment about $\in$ in the rule?} \textbf{Unfold} increments the bound on the recursion depth.
%\textbf{Safe} returns $\may_i$ as invariant when the sequence of may summaries
%converges. \textbf{Unsafe} applies when the must summary intersects with $\Bad$.
The other rules are similar to IC3~\cite{ic3} and GPDR~\cite{gpdr} and we skip their explanation in the interest of
space. \spacer is sound and if \MBP utilizes finite MBP functions, \spacer also
terminates for a fixed $N$~\cite{spacer_cav14}.

%One can show that $\Init \implies \may_0$ and for a fixed $N$ and every $0 < i \le
%N$, $\cF(\may_{i-1}) \implies \may_i$, $\may_{i-1} \implies \may_i$, and $\may_i
%\implies \neg\Bad$. Moreover, $\cF^i(\Init) \implies \may_i$ for all $i$, and
%$\must \implies \cF^N(\Init)$. Thus, $\{\may_i\}_{i \in \nat}$ and $\must$,
%respectively, over- and under-approximate reachable states and \spacer is sound.

%In the description above, we left out many implementation details and
%we mention a few of them here. For efficiency, we restrict queries to
%cubes. For Linear Arithmetic, we use \MBP functions that are
%linear in time and space.  $\Queue$ is maintained as a priority queue,
%processing queries of smaller recursion depths first. Additional
%constraints are imposed on the rules and their ordering to ensure
%termination for a fixed $N$~\cite{spacer_cav14}. For the rule \textbf{Unsafe}, our implementation also
%produces a counterexample in addition to returning \textsc{Unsafe}.

\subsection{Instantiation for ARR+LIA}

In instantiating this framework for ARR+LIA, the key ingredient is the
\textsc{MBP} procedure of the previous section. An interpolation
procedure \textsc{Itp} can be trivially obtained by using
literal-dropping approach based on UNSAT cores, or a more sophisticated
approach can be taken (e.g., see~\cite{gpdr,duality}).

Because we do not have a \emph{finite} MBP, \spacer is not guaranteed
to terminate even for a fixed bound on the recursion depth $N$. That
is, it can generate an infinite sequence of queries and must summaries.
Note that MBP is used in 3 rules: \textbf{DecideMay},
\textbf{DecideMust}, and \textbf{Successor}. The elimination of
quantifiers in \textbf{Successor} is only an optimization and can be
avoided. This is not the case with \textbf{DecideMay} or
\textbf{DecideMust} without changing the structure of the queries, the
considerations of which are outside the scope of this paper. In the
following, we identify restrictions on the CHCs where termination is
still guaranteed and for the other cases, we propose some heuristic
modifications to \textsc{Mbp} and \textsc{Itp} to help avoid
divergence.

\subsubsection{Equality resolution in \textsc{Mbp}}
There are several cases where terms over combined signatures appear in
conjunction with equality terms over the index quantifier, e.g., $\exists i:\int
\st i=t \land \rd{a}{i} > 0$ for a term $t$ independent of $i$. In these cases, the
quantifier can be eliminated using equality resolution, e.g., $\rd{a}{t} > 0$ in
the above example. Such cases seem to be natural in the case of a single
procedure, i.e., when $\Tr$ in (\ref{eq:safety}) is independent of $\xcall$. Consider
a disjunct $\delta$ in a DNF representation of $\Tr$. Now, $\delta$ represents a
path in the procedure and typically, index terms (in reads and writes) in
$\delta$ can be ordered such that every index term is a function of the previous
index terms or the current-state variables $\vec{x}$. This makes it possible to
eliminate any index variables in $\vec{x}'$ using equality resolution as
mentioned above.

%So, given $\exists i : \int \st \varphi$ and $M \models \varphi$, if $i$
%appears inside any array term in $\varphi$, we under-approximate the
%quantification by substituting $i$ throughout $\varphi$ using $M(i)$. In the
%above example, given $M \models \rd{a}{i} > 0$ with $M(i) = 2$, we
%under-approximate it as $\rd{a}{2} > 0$. \spacer is no longer guaranteed to
%terminate even for a fixed bound on the recursion depth ($N$), as the following
%set of CHCs shows.

\subsubsection{Privileging array equalities}

Here is a simple example that exhibits non-termination:
\begin{align*}
    \Inv(a,b) &\gets a=b \\
         \bot &\gets \Inv(a,b), \rd{a}{j} < 0, \rd{b}{j} > 0
\end{align*}
Here, intuitively, $\Inv(a,b)$ denotes the summary of a procedure which takes
an array $a$ as input and produces $b$ as output and we are interested in checking if
there is sign change in the value at an index $j$ as a result of the procedure
call. For this example,
\textbf{DecideMay} creates queries of the form $\rd{a}{k} < 0 \land \rd{b}{k} >
0$ where $k$ is a specific integer constant. If $\Itp$ returns interpolants of
the form $\rd{a}{k} = \rd{b}{k}$, it is easy to see that \spacer would not
terminate even for $N = 0$, even though there is a trivial solution: $a = b$.

To alleviate this problem, we modify \textsc{Mbp} and \textsc{Itp} to promote the
use of array equalities in interpolants. Let $\psi$ be the result of \MBP
for a given model $M$. For every pair of array terms $a$, $b$ in
$\psi$, we strengthen $\psi$ with the array equality $a=b$ or disequality $a
\neq b$, depending on whether $M \models a=b$ holds or not. In the above
example, the queries will now be of the form $\rd{a}{k} < 0 \land \rd{b}{k} > 0
\land a \neq b$. However, $\rd{a}{k} = \rd{b}{k}$ continues to be an interpolant
whereas the desired interpolant is $a = b$. To reduce the dependence on specific
integer constants in the learned interpolants, and hence in the may summaries, we
modify \Itp as follows. Suppose we are computing an interpolant for $\psi \implies \neg\varphi'$ (as
occurs in \textbf{Conflict}). We let $\varphi = \varphi_1 \land \varphi_2$ where
$\varphi_2$ contains all the literals where an integer quantifier is substituted
using its interpretation in a model. Using a \emph{minimal unsatisfiable
subset} (MUS) algorithm, we can generalize $\varphi_2$ to $\hat{\varphi}_2$ such that
$\psi \land \left(\varphi_1 \land \hat{\varphi}_2\right)'$ is unsatisfiable and then
obtain $\Itp(\psi, \neg \left(\varphi_1 \land \hat{\varphi}_2\right)')$. In the above
example, for $N=0$ we have $\psi = (a=b)$, $\varphi_1 = (a \neq b)$, and
$\varphi_2 = \rd{a}{k} < 0 \land \rd{b}{k} > 0$. One can show that
$\hat{\varphi}_2$ is simply $\top$ and the only possible interpolant is $a=b$.
In our implementation, we add such (dis-)equalities on-demand in a lazy fashion.
Note that adding such (dis-)equalities to the queries is only a heuristic and
may not always help with termination.

%%% Local Variables:
%%% mode: latex
%%% TeX-master: "main"
%%% End:

\section{Experimental Results}
\label{sec:results}

As noted in the introduction, the array theory allows us to model heap
references accurately. This eliminates the need to inline procedures
so that heap-allocated objects are reduced to local
variables. We hypothesize that the resulting increase in modularity
will allow \spacer to more efficiently verify procedural programs using
\textsc{ArrayMbp}, in spite of the potential for divergence due to
non-finiteness of the MBP.

We test this hypothesis using a prototype implementation of \spacer
with
\textsc{ArrayMbp}.\footnote{\url{https://bitbucket.org/spacer/code}}
To verify C programs, we use \seahorn~\cite{seahorn}, which uses the
LLVM infrastructure to compile and optimize the input program, then
encodes the verification conditions as CHCs in the SMT-LIB2
format. 
\seahorn can optionally inline procedure calls before
encoding, allowing us to test our hypothesis regarding modularity.
% To test our hypothesis, we inline procedures using an unfolding
% transformation on the CHC representation.

For reference, we also compare \spacer to the implementation of
GPDR~\cite{gpdr} in Z3~\cite{z3}.  A key difference between \spacer
and GPDR is that the latter does not use must summaries. Z3 also uses
MBP, but is limited to equality resolution and the substitution
method. As a result Z3 GPDR is effective only for inlined programs.

We use benchmarks from the software verification competition
SVCOMP'15~\cite{svcomp15}.  We considered the 215 benchmarks from the
\emph{Device Drivers} category where Z3 GPDR (with inlining) needed more than a minute
of runtime or did not terminate within the resource limits of
SVCOMP~\cite{seahorn_svcomp15}.
All experiments have been carried out using a 2.2 GHz AMD Opteron(TM)
Processor 6174 and 516GB RAM, running Ubuntu Linux. Our resource
limits are 30 minutes and 15GB for each verification task. In the
scatter plots that follow, a diamond indicates a \emph{time-out}, a
star indicates a \emph{mem-out}, and a box indicates an anomaly in the
implementation.

\begin{figure}[t]
\centering
\includegraphics[scale=.8]{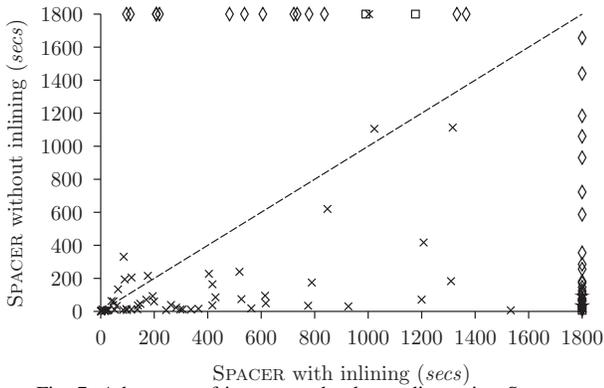}
\vspace{-0.1in}
\caption{Advantage of inter-procedural encoding using \spacer.}
\label{fig:spacer_inl_noinl}
\vspace{-0.1in}
\end{figure}

\begin{figure}[t]
%\begin{subfigure}[b]{.5\textwidth}
\centering
\includegraphics[scale=.8]{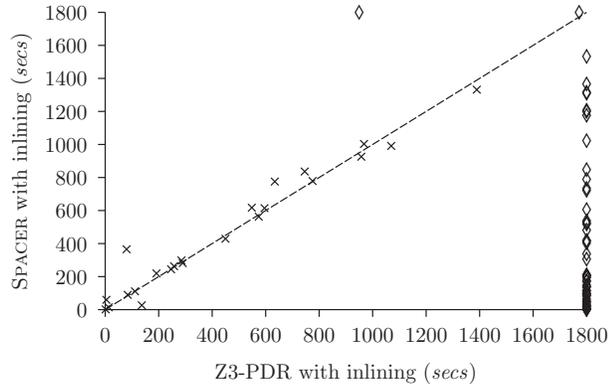}
%\caption{}
%\end{subfigure}
% \begin{subfigure}[b]{.5\textwidth}
% \centering
% \includegraphics[scale=.8]{data/spacer_vs_pdr_smt2/plot}
% \caption{}
% \label{fig:spacer_pdr_noinl}
% \end{subfigure}
% \caption{\spacer vs. Z3 on hard benchmarks (a) with and (b) without inlining}
\caption{\spacer vs. Z3 on hard SVCOMP benchmarks with inlining.}
\label{fig:spacer_pdr_inl}
\vspace{-0.1in}
\end{figure}

The scatter plot in Fig.~\ref{fig:spacer_inl_noinl} compares the
combined run time for the CHC encoding and verification, when inlining
is turned on and off. A clear advantage is seen in the
non-inlining case. This shows that \spacer is able to effectively
exploit the additional modularity that is made possible by \textsc{ArrayMBP},
and that this advantage outweighs any occurrences of divergence due to
non-finite MBP.\footnote{Unfortunately, we have no way to distinguish divergence from timeouts.}
We note that \spacer with only LIA is able to handle only a small fraction of the non-inlined benchmarks.
This result confirms our hypothesis.

For reference, we also compare to the performance of Z3 GPDR. We observed that
without \textsc{ArrayMBP}, Z3 is very ineffective in the non-inlined case.
We should mention, however, that of the 7 unsafe programs verified by Z3, 5 could
not be verified by \spacer.
Fig.~\ref{fig:spacer_pdr_inl} compares \spacer and Z3 with inlining
on. This shows an overwhelming advantage for \spacer, which is due to
its more effective MBP approach.

% \textbf{(b) Advantage of our compositional framework}.
% To see the effect of must summaries and MBP (for LIA and ARR), we compared Z3
% and \spacer. The scatter plots in Fig.~\ref{fig:spacer_pdr_inl}
% and~\ref{fig:spacer_pdr_noinl} compare the tools on the CHCs obtained when inlining in \seahorn is
% turned on and off, respectively. In both cases, we clearly see that \spacer has
% a significant advantage. Note that, in the latter case, Z3 runs out of time on
% most of the benchmarks verifying 10 programs (3 safe; 7 unsafe) while \spacer
% can verify 97 programs (21 safe; 76 unsafe). 

% In summary, we believe \spacer is a valuable addition to the state-of-the-art as
% shown above by its practical advantage on some hard device driver benchmarks.

\section{Related Work}
There are several SMT-based approaches for sequential program verification that
iteratively check satisfiability of formulas corresponding to safety of various
unwindings of the program~\cite{whale,hsf,ultimate,duality}. However, these
monolithic SMT formulas can grow exponentially. In contrast, the \spacer
framework~\cite{spacer_cav14} we use allows us to do a compositional proof
search for safety. Such local proof search is also found in the IC3 algorithm
for hardware model checking~\cite{ic3} and its extensions to software model
checking (e.g.,~\cite{gpdr}), although \spacer is the first to use
under-approximate summaries of procedures for avoiding redundant proof
sub-goals.
%  Moreover, given the infinite size of the state-space in program
% verification, a finite MBP is critical for ensuring termination of the
% compositional proof search, which is also unique to \spacer.
Model-based
generalizations have also been used to obtain projections efficiently
in decision procedures for quantified formulas~\cite{lazy_qe}.

%arrays (restrictions exist: array property fragment, EPR. But we go unrestricted in the theory but can only generate qfree proofs)

\section{Conclusion and Future Work}
%quantified: safari, bmr sas?, smt-based array invariant?
We have presented a procedure for existentially projecting array
variables from formulas over combined theories of ARR, LIA, and
propositional logic. We have adapted the procedure to a finite
MBP for array variables. While existential projection is worst-case exponential, the
corresponding MBP is polynomial. However, projecting arrays might
introduce new existentially quantified variables (whose sort is the
same as the index- or value-sort of the eliminated array). For
projecting these variables, a finite MBP need not exist. We described
heuristics for obtaining a practical (but not necessarily finite) MBP
procedure, obtaining an instantiation of the \spacer framework for
verification of safety of sequential heap-manipulating programs. We
show that the new variant of \spacer is effective for constructing
compositional proofs of Linux Device Drivers. In the future, we plan to
extend these ideas for handling more complex heap-manipulating
programs that require universal quantifiers in the program invariants.

%%% Local Variables:
%%% mode: latex
%%% TeX-master: "main"
%%% End:

% Generated by IEEEtranS.bst, version: 1.12 (2007/01/11)

\newpage
\appendices

\section{QE and MBP for ARR over Finite Index Domains}
\label{app:finite_dom}

\nsb{a brute force approach could be that you simply introduce fresh variables
$v_1, \ldots, v_k$ corresponding to the size of the domain. Then for every
occurrence of rd(a,j) insert the term $ite(j=1,v_1,ite(j=2,v_2,\ldots))$
(or the context expanded version as you prefer), and for every
equality $a =_i t$ a conjunction $(v_1 = rd(1) \lor 1 \in i), \ldots$.
If the size of the index domain is larger than the number of subterms
of the original formula you don't need to apply this transformation and 
can just leverage a ``large'' domain assumption to introduce enough
distinct indices to satisfy disequalities between subterms.
}
When finite interpretations of $I$ are allowed, \textsc{ElimDiseq} is no longer
an equivalent transformation as there may not exist an index where the arrays in
the disequalities disagree on the values. However, one can use extensionality to
obtain another equivalent transformation rule \textsc{ElimDiseqFinite}, as shown
in Fig.~\ref{fig:diseq_finite}. As this rule introduces new read terms over
$a$, we need to apply \textsc{FactorRd} once again before \textsc{Ackermann}. Also, note that the result of QE and MBP is now of the form
$\exists \idx : I, \val : V \st \psi$.

\begin{figure}[t]
\begin{mathpar}
    \inferrule*[left=ElimDiseqFinite]
    {\exists a \st \left( \neg (\peq{a}{t}{\idx}) \land \varphi \right)}
    {\exists a,j \st \left( \rd{a}{j} \neq \rd{t}{j} \land j \not\in \idx \land \varphi \right)}
\end{mathpar}
\flushright{where $a$ does not appear in $t$}
\caption{Modified version of \textsc{ElimDiseq} for finite domains.}
\label{fig:diseq_finite}
\end{figure}

\section{Proofs of statements about \textsc{ArrayQE} and \textsc{ArrayMBP}}
\label{app:qe_proofs}

\newcounter{oldtheorem}

\setcounter{oldtheorem}{\thetheorem}
\setcounter{theorem}{0}
\begin{theorem}
    $\textsc{ArrayQE}(\exists a : \arr{I}{V} \st \varphi)$ returns $\exists \val
    : V \st \rho$, where $\rho$ is quantifier-free and $\exists \val \st \rho
    \equiv \exists a \st \varphi$.
\end{theorem}
\begin{proof} (\emph{Sketch})
    One can easily show that the rules in Fig.~\ref{fig:str_rules},
    \ref{fig:eq_sel_rules}, and~\ref{fig:core_qe_rules} are equivalence
    preserving. The theorem follows immediately.
\end{proof}
\setcounter{theorem}{\theoldtheorem}

\setcounter{oldtheorem}{\thetheorem}
\setcounter{theorem}{1}
\begin{theorem}
    $\textsc{ArrayQE}(\exists a \st \varphi)$ terminates in time exponential in
    the size of $\varphi$.
\end{theorem}
\begin{proof} (\emph{Sketch})
    Line~1 of \textsc{ArrayQE} essentially eliminates write terms one by one and
    can be easily shown to terminate. Line~2 can be easily made to terminate by
    iterating over all partial equality and read terms. The remaining steps of
    the algorithm clearly terminate as well.

    The complexity analysis is similar to that of the decision procedure by Stump et
    al.~\cite{stump}. Let $N$ be the size of $\varphi$. The number of disjuncts
    generated by any rewrite rule is bounded by $N$ (due to the disjunction $j
    \in \idx$ on indices in \textsc{ElimWrEq}). Disjunctions can be
    generated by the rules for every write term or partial equality and their
    number is bounded by $N$.  So, the total number of disjunctions generated by
    the algorithm is bounded by $O(N^N)$ which is exponential in $N$.  The size
    of a disjunct generated by a rule can be shown to be bounded by a polynomial
    in $N$. \textsc{CaseSplitEq} can be efficiently implemented using an $(N+1)$-way
    case analysis over all $N$ partial equalities at once avoiding a Boolean rewriting on line~3 of
    the algorithm. That is, one can obtain $N+1$ disjuncts, one each for the
    case of a partial equality being true and the last one for the case of every
    partial equality being false. Thus, the complexity of \textsc{ArrayQE} is exponential in
    $N$.
\end{proof}
\setcounter{theorem}{\theoldtheorem}

\end{document}